\documentclass[aps,prb,amsmath,amssymb,twocolumn,showpacs,showkeys,superscriptaddress]{revtex4-1}

\usepackage{graphicx}
\usepackage{dcolumn}
\usepackage{color}
\usepackage{ulem}

\begin{document}
\title{Effect of second order piezoelectricity on excitonic structure of stress-tuned InGaAs/GaAs quantum dots}

\author{Petr Klenovsk\'y}
\email[]{klenovsky@physics.muni.cz}
\affiliation{Department of Condensed Matter Physics, Faculty of Science, Masaryk University, Kotl\'a\v{r}sk\'a~267/2, 61137~Brno, Czech~Republic}
\affiliation{Central European Institute of Technology, Masaryk University, Kamenice 753/5, 62500~Brno, Czech~Republic}

\author{Petr Steindl}
\affiliation{Department of Condensed Matter Physics, Faculty of Science, Masaryk University, Kotl\'a\v{r}sk\'a~267/2, 61137~Brno, Czech~Republic}
\affiliation{Central European Institute of Technology, Masaryk University, Kamenice 753/5, 62500~Brno, Czech~Republic}

\author{Johannes Aberl}
\affiliation{Institute of Semiconductor and Solid State Physics, Johannes Kepler University Linz, Altenbergerstra{\ss}e 69, A-4040 Linz, Austria}

\author{Eugenio Zallo}
\affiliation{Institute for Integrative Nanosciences, IFW Dresden, Helmholtzstra{\ss}e 20, D-01069 Dresden, Germany}
\affiliation{Paul-Drude-Institut f{\"u}r Festk{\"o}rperelektronik, Hausvogteilplatz 5-7, 10117 Berlin, Germany}



\author{Rinaldo Trotta}
\affiliation{Institute of Semiconductor and Solid State Physics, Johannes Kepler University Linz, Altenbergerstra{\ss}e 69, A-4040 Linz, Austria}
\affiliation{Department of Physics, Sapienza University of Rome,
Piazzale Aldo Moro 5, 00185 Rome, Italy}

\author{Armando Rastelli}
\affiliation{Institute of Semiconductor and Solid State Physics, Johannes Kepler University Linz, Altenbergerstra{\ss}e 69, A-4040 Linz, Austria}

\author{Thomas Fromherz}
\affiliation{Institute of Semiconductor and Solid State Physics, Johannes Kepler University Linz, Altenbergerstra{\ss}e 69, A-4040 Linz, Austria}

\date{\today}

\begin{abstract}

We study the effects of the nonlinear piezoelectricity and the In distribution on the exciton energy, the electron-hole electric dipole moment, and the fine-structure splitting in stress-tunable InGaAs/GaAs quantum dots integrated onto a piezoelectric actuator. In particular, we investigate in detail the contributions of various elements of the expansion of the electrical polarization in terms of externally induced elastic strain on the latter two important quantum dot properties. Based on the comparison of the effects of first- and second-order piezoelectricity we provide a simple relation to estimate the influence of applied anisotropic stress on the quantum dot dipole moment for quantum dots significantly lattice mismatched to the host crystal. 

\end{abstract}

%

\maketitle


Semiconductor Quantum dots (QDs) provide a number of appealing applications. Among others, QDs may be used as gain materials for lasers, \cite{Bimberg1997,Ledentsov,Heinrichsdorff1997} as single photon emitters for optical fibre communication,~\cite{Huffaker1998} as building blocks of secure optical links using entangled photon pairs,~\cite{Trotta:16} for quantum gates,~\cite{Krapek2010,Klenovsky2016} or are used in biomedical applications.~\cite{Wegner2015}
%

The capability of QDs to confine the motion of electrons and holes in all three spatial dimensions offers the advantages of a discrete, atom-like electronic system~\cite{kastner1993artificial} within a solid-state platform. The strong confinement and the Coulomb interaction among trapped charge carriers promotes the formation of stable few-particle states~\cite{PhysRevB.79.075443} like neutral exciton ($X$) and biexciton ($XX$) whose cascaded radiative recombination allows the generation of single- and entangled photons.~\cite{michler2000quantum,PhysRevLett.84.2513} The application of QDs as quantum light sources in advanced quantum communication and computation schemes~\cite{senellart2017high,huber2018semiconductor} demands well-defined transition energies, vanishing fine-structure splitting (FSS) and extensive control over the QDs' interaction with the charge environment. In this regard, the statistical distribution of structural parameters such as size, shape, or composition of self-assembled QDs~\cite{PhysRevB.76.205324}, which becomes apparent via deviations of essential emission properties among different QDs, represents a major challenge towards application and demands for effective methods for (reversible) post-growth engineering~\cite{Plumhof2012} of the electronic structure of individual QDs.

In this context, externally applied stress mediated via piezoelectric actuators~\cite{Martin-Sanchez2018} has proven to be an effective tool to (simultaneously) tune transition energies~\cite{PhysRevB.88.155312} and FSS~\cite{PhysRevLett.109.147401,Trotta:16}, thus, enabled interference experiments with photons from remote QDs~\cite{doi:10.1021/acs.nanolett.7b00777} or the extraction of high-fidelity polarization-entangled photons.~\cite{huber2018strain} In a recent work~\cite{Aberl:17} we moreover demonstrated that applied stress allows to control magnitude and alignment of the vertical electron-hole separation in In(Ga)As QDs manifesting itself via a built-in dipole moment ($p$) along the growth direction. $p$ is commonly present in as-grown QDs \cite{Fry:00} and its interaction with charges in their vicinity leads to spectral diffusion~\cite{Callsen:15} causing an inhomogeneous linewidth broadening of the corresponding optical transitions and, in turn, degrade the indistinguishability of consecutive photons emitted by QD. It has been found in Ref.~[\onlinecite{Aberl:17}] that the observed tuning of $p$ can only be described by considering nonlinear terms in the expansion of the piezoelectric polarization, the importance of which was first highlighted theoretically by Bester \textit{et al.}.~\citep{Bester:06, Bester:06_2} However, that effect is usually difficult to observe experimentally.

In this work we discuss the significance of the second-order piezoelectric terms with regard to the FSS in stress-tuned In(Ga)As QDs. In addition, the previously reported dependences of the $X$ transition energy ($E_0$) and $p$ on the externally applied stress~\cite{Aberl:17} are analyzed in more detail. The experimental data in Ref.~[\onlinecite{Aberl:17}] were obtained on QDs embedded in {n-i-p} membrane-diodes bonded on a PMN-PT piezoelectric actuator. This device design allowed to extract $E_0$ and $p$ vs. applied stress from micro-photoluminescence ($\mu$-PL) measurements of the quantum-confined Stark effect (QCSE)~\cite{PhysRevB.70.201308} whereas the corresponding FSS was obtained via polarization-resolved $\mu$-PL measurements of the $X$ and $XX$ spectral lines. The presented theoretical model allows to concurrently reproduce the experimental data for the considered quantities in terms of magnitude and observed stress-dependence. This is achieved while using realistic structural parameters for the investigated QDs and taking into account peculiarities of the used device ({-processing}) in terms of stress-configuration and prestress. The performed analysis finally allows us to propose an approximate relation of $p$ and the externally applied stress applicable to all epitaxial QD systems lattice mismatched with the host material. We want to emphasize that this analysis is not only applicable to type-I QD systems like InGaAs/GaAs, but also for QD systems supporting spatial indirectly located electron and hole states (type-II QDs) that have been reported for distinct $\text{III--V}$ material combinations.~\cite{Klenovsky2015,Klenovsky2017} 

%

The Taylor expansion of the electrical polarization (${\bf P}$) in terms of strain ($\eta$) up to second-order terms is ${\bf P}={\bf P}_{l}+{\bf P}_{nl}$,~\cite{Beya-Wakata2011} where ${\bf P}_{l}$ is the linear term:
%
%
%
\begin{equation}
\label{eq:1stPiez}
{\bf P}_{l}=e_{14}\begin{pmatrix}\eta_4\\\eta_5\\\eta_6\end{pmatrix},
\end{equation}
and ${\bf P}_{nl}$ the nonlinear one:
%
%
\begin{equation}
\label{eq:2ndPiez}
{\bf P}_{nl}=B_{114}\begin{pmatrix}\eta_1\eta_4\\\eta_2\eta_5\\\eta_3\eta_6\end{pmatrix}+
B_{124}\begin{pmatrix}\eta_4(\eta_2+\eta_3)\\\eta_5(\eta_3+\eta_1)\\\eta_6(\eta_1+\eta_2)\end{pmatrix}+
B_{156}\begin{pmatrix}\eta_5\eta_6\\\eta_4\eta_6\\\eta_4\eta_5\end{pmatrix}.
\end{equation}
Here $\eta_i$ are indexed according to the Voigt notation,~i.e.,~ $\eta_1\equiv\eta_{xx}$, $\eta_2\equiv\eta_{yy}$, $\eta_3\equiv\eta_{zz}$, $\eta_4\equiv2\eta_{yz}$, $\eta_5\equiv2\eta_{xz}$, $\eta_6\equiv2\eta_{xy}$,~\cite{Beya-Wakata2011} where $x,y,z$ denote the crystallographic axes of the conventional cubic unit cell of the zincblende lattice.
%
%
Note that even though the third order coefficients of the above expansion were provided by Tse and colleagues,~\cite{Tse2013} we restrict ourselves to second-order ones in this work since the magnitude of externally induced (missfit) $\eta$ is of the order 0.1\,\% (3\,\%).~\cite{Aberl:17} As a consequence, the largest third-order contributions involving the externally induced strain are products of that with the squared misfit strain. These contributions are much smaller than the largest second-order contributions involving the misfit strain in first order. 


In the simulations discussed in this work the calculation flow was as follows. First, the geometry of the QD structure was defined on a rectangular grid including the spatially dependent material constituents. Thereafter, the strain field in and around QD was found by minimizing the strain energy. The effect of the resulting strain on the confinement potential was then treated using the Bir-Pikus hamiltonian~\cite{Bir:74} with positionally dependent parameters. The next step involved the self-consistent solution of single-particle Schr\"{o}dinger and Poisson equations including the effect of piezoelectric fields up to the second-order in $\eta$. Note that the single-particle states were obtained within the envelope function method based on {8-band} $\mathrm{{\bf k}}\cdot{ \mathrm{\bf p}}$ approximation and all the preceding steps of calculation were done using the nextnano$3$ simulation suite.~\cite{Birner:07} For the full list of material parameters used in this work see Ref.~[\onlinecite{SupMatParam}]. Finally, the obtained single-particle states were used as input for the excitonic calculations using the Configuration Interaction (CI) algorithm that we have previously developed.~\cite{Klenovsky2017} All CI calculations included the computation of direct and exchange Coulomb integrals and were performed with a basis-set of six electron and six hole single-particle states, thus, providing also the effect of correlation.

%
%


\begin{figure}
\renewcommand{\tabcolsep}{2pt}
\begin{center}
\begin{tabular}{c}
\includegraphics[width=0.45\textwidth]{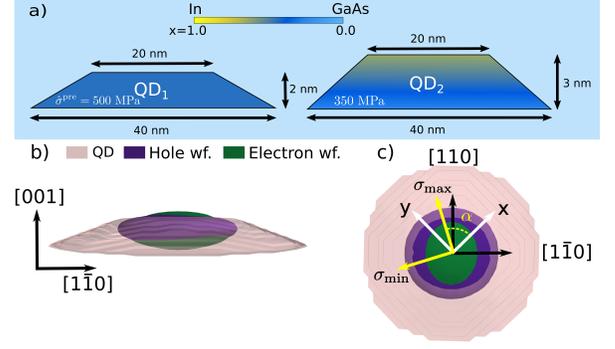} \\ 
\end{tabular}
\end{center}
\caption{a) Side view of the In$_{{x}}$Ga$_{1-x}$As/GaAs QD$_1$ and QD$_2$ structures used for the calculations. The shape of both QDs is that of truncated cones with base and top diameters of 40\,nm and 20\,nm, respectively. The height is 2~nm~(3~nm), the In concentration is equal to 0.45 (linearly increasing from 0.25 at the bottom to 0.65 at the apex), and $\sigma^\text{pre}=500$\,MPa ($\sigma^\text{pre}=350$\,MPa) for QD$_1$ (QD$_2$). b) side and c) top view of the typical simulated dot (pink), and calculated electron (green) and hole (blue) probability densities, respectively. The wavefunctions are given as isosurfaces encircling 70\,\% of the probability.
\label{fig:QDStruct}}
\end{figure}

\begin{figure}[!ht]
\renewcommand{\tabcolsep}{2pt}
\begin{center}
\begin{tabular}{c}
\includegraphics[width=0.45\textwidth]{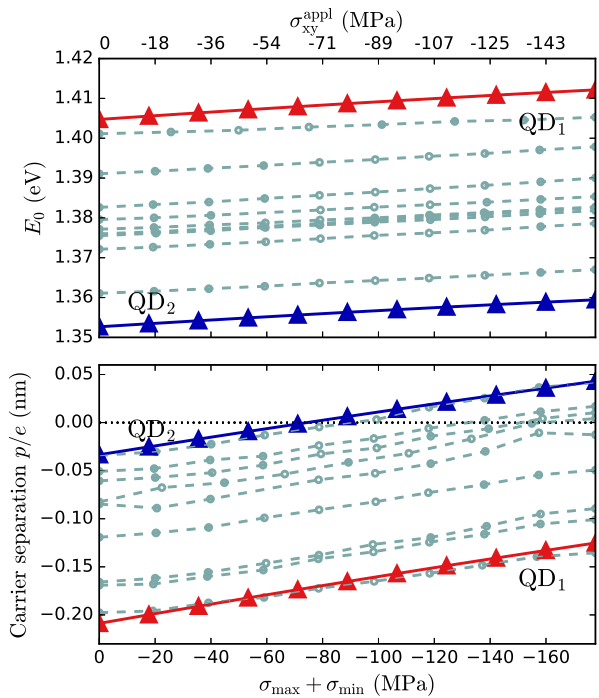}\\
\end{tabular}
\end{center}
\caption{
Dependencies of average energy $E_0$ (top panel) and average electron-hole separation $p/e$ (bottom panel) on $\sigma_{\mathrm{max}}+\sigma_{\mathrm{min}}$ experimentally obtained from $\mu$PL measurements of nine InGaAs QDs~\cite{Aberl:17} (broken curves) and that calculated for QD$_1$ (full red curve) and QD$_2$ (full blue curve). The different upper and lower x-scales uniquely define the in-plane applied stress tensor via the relation~\cite{SupMatPrincStr} $\sigma_{xy}^\text{appl}$=$\frac{1}{2}(\sigma_\text{max}-\sigma_\text{min})\sin(2\alpha)$, where $\alpha$ is given in the text. The letter $e$ denotes the elementary charge.  
\label{fig:TheorVsExp}}
\end{figure}

Two In$_{x}$Ga$_{1-x}$As/GaAs QDs labeled QD$_1$ and QD$_2$ in Fig.~\ref{fig:QDStruct}~a) were used as model systems. Both have the shape of truncated cones but differ in size and In-Ga alloy distribution. Their parameters were deliberately chosen so that the calculated dependencies of $E_0$ and $p$ on the hydrostatic part of the applied anisotropic stress $\sigma_{\mathrm{max}}+\sigma_{\mathrm{min}}$ match the experimental results taken from Ref.~[\onlinecite{Aberl:17}], see Fig.~\ref{fig:TheorVsExp}. Note, that similarly as in Ref.~[\onlinecite{Aberl:17}] $p$ is considered to be oriented from negative to positive charge throughout this work.
The variables $\sigma_{\mathrm{max}}$ and $\sigma_{\mathrm{min}}$ denote the principal stresses~\cite{Trotta:15,SupMatPrincStr} applied externally by the two-dimensional piezo actuator. In  Ref.~[\onlinecite{Aberl:17}] it was shown that $\sigma_{\mathrm{max}}$ was applied at an angle of $\alpha=55^{\circ}$ with respect to the [100] crystal axis which we adopt also in this work. This stress configuration corresponds to the experimental one as estimated via the measurements of FSS, see supplementary of Ref.~[\onlinecite{Aberl:17}]. The various coordinate systems used in our model as well as the typical single-particle wavefunctions of electrons and holes are indicated in Fig. \ref{fig:QDStruct} b) and c).
Note that by assuming a smaller average In concentration (45\,\% instead of 62.5\,\%) but a larger In gradient along growth direction (from 25\,\% to 65\,\% instead of 45\,\% to 80\,\%) as compared to  Ref.~[\onlinecite{Aberl:17}], in this work we could significantly improve the agreement between simulated and measured slope of $p/e$ with applied stress. At the same time, the observed scattering range of $X$ energies and dipole moments remains within a model parameter region of comparable width as assumed in Ref.~[\onlinecite{Aberl:17}].
%
%
%
%

\begin{figure}[!ht]
\renewcommand{\tabcolsep}{2pt}
\begin{center}
\begin{tabular}{c}
\includegraphics[width=0.45\textwidth]{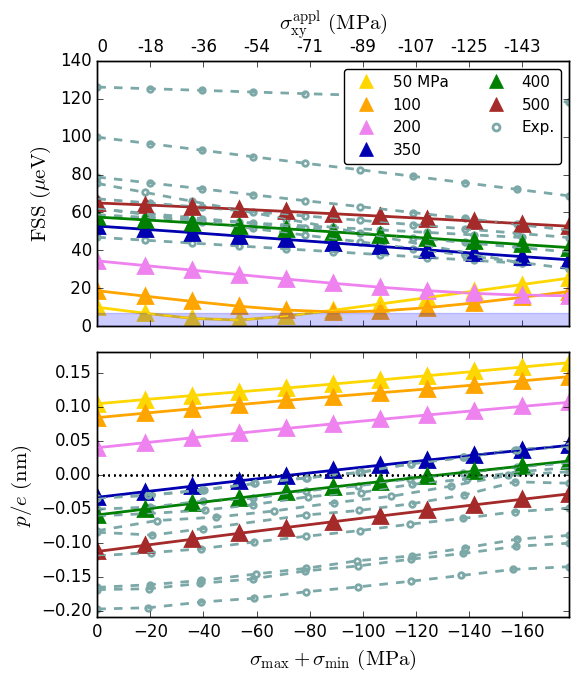} \\
\end{tabular}
\end{center}
\caption{
Dependencies of FSS (top panel) and $p/e$ (bottom panel) on $\sigma_{\mathrm{max}}+\sigma_{\mathrm{min}}$ experimentally obtained from $\mu$PL measurements of nine InGaAs QDs~\cite{Aberl:17} (broken curves) and that calculated for different values of $\sigma^{\mathrm{pre}}_{xy}$ as indicated in the legend. Except for $\sigma^{\mathrm{pre}}_{xy}$ the simulated QDs had the same properties as QD$_2$. The shaded area in the top panel indicates the range of FSS variations due to dot elongation along $[1\overline{1}0]$ crystallographic direction in the range between 0.9 to 1.2. The meaning of both x-scales is the same as in Fig.~\ref{fig:TheorVsExp}.
\label{fig:TuningByPrestress}}
\end{figure}
As discussed in Ref.~[\onlinecite{Aberl:17}], the bonding of the sample onto the piezo actuator leads to a prestress ($\sigma^\text{pre}$) independent on the voltage applied to the piezo varying between different dots. As will be discussed towards the end of the paper, only the off-diagonal component  $\sigma_{xy}^\text{pre}$ of the (symmetric) in-plane prestress tensor effectively affects the electron-hole separation $p/e$ in QDs, where $e$ denotes the elementary charge. Consequently, in order to match the measured values of $p/e$ with the results of our calculations we needed to allow for different magnitudes of $\sigma_{xy}^\text{pre}$ of 500 and 350\,MPa that acted on QD$_1$ and QD$_2$, respectively. We support this assumption by comparing measured values of FSS with those obtained using CI, as discussed in the following.

The effects of $\sigma_{xy}^\text{pre}$ on FSS and $p/e$ are different, however, it is possible to estimate a value of $\sigma_{xy}^\text{pre}$ such that one can fit both sets of experimental data, i.e., for FSS and $p/e$.
In the top panel of Fig.~\ref{fig:TuningByPrestress} we show that for QD$_2$ the application of a variable stress
%
%
%
leads to a minimal FSS of 1.15\,$\mu$eV for $\sigma_{\mathrm{max}}+\sigma_{\mathrm{min}}=-53.32$\,MPa if we assume $\sigma_{xy}^\text{pre}=50$\,MPa. Note that by the two scales of the abscissa axes in Fig.~\ref{fig:TuningByPrestress} together with $\alpha = 55^\circ$ and the relation~\cite{SupMatPrincStr} $\sigma_{xy}^\text{appl} = \frac{1}{2}(\sigma_\text{max} - \sigma_\text{min})\sin(2\alpha)$ all components of the in-plane stress tensor are defined.  
For larger values of  $\sigma_{xy}^\text{pre}$ the applied stress leads again to reduction of FSS, but the minimal value of FSS is progressively larger as well as the value of $\sigma_{\mathrm{max}}+\sigma_{\mathrm{min}}$ for which the anticrossing occurs. 
At the same time, the values of $p/e$ for ${\sigma_{\mathrm{max}}+\sigma_{\mathrm{min}}=0}$ decrease with increasing  $\sigma_{xy}^\text{pre}$, see bottom panel of Fig.~\ref{fig:TuningByPrestress}. Interestingly, $p/e$ attains positive values for  $\sigma_{xy}^\text{pre}\lesssim 200$\,MPa. However, larger values of $\sigma_{xy}^\text{pre}$ lead to negative values of $p/e$ for $\sigma_{\mathrm{max}}+\sigma_{\mathrm{min}}=0$. Notice that $\partial p/\partial(\sigma_{\mathrm{max}}+\sigma_{\mathrm{min}})$ is very similar among different dots. We will return to discussion of this observation later.

It is well known, that apart from $\sigma_{xy}^\text{pre}$ FSS also depends on the elongation of the QDs along $[1\overline{1}0]$ crystallographic direction.~\cite{Zielinski2013,Singh2018} However, our simulations show that for large QDs with dimensions similar to that of QD$_2$, such elongations in an unrealistically large range between 0.9 and 1.2 cause FSS of less than $\sim10\,\mu$eV, see Ref.~[\onlinecite{SupMatElong}]. Since the elongation-induced FSS  is much smaller than the FSS observed in our experiments, it was neglected in our analysis.
%

Our model reproduces the experimental values of FSS and $p/e$ as well as $\partial \text{FSS}/\partial(\sigma_{\mathrm{max}}+\sigma_{\mathrm{min}})$ and $1/e\times\partial p/\partial(\sigma_{\mathrm{max}}+\sigma_{\mathrm{min}})$ reasonably well for $\sigma_{xy}^\text{pre}\gtrsim 350\,$MPa indicating that rather large  $\sigma_{xy}^\text{pre}$ is experienced by our QDs and the value of that is different among dots.

\begin{figure}[!ht]
\renewcommand{\tabcolsep}{2pt}
\begin{center}
\begin{tabular}{c}
\includegraphics[width=0.45\textwidth]{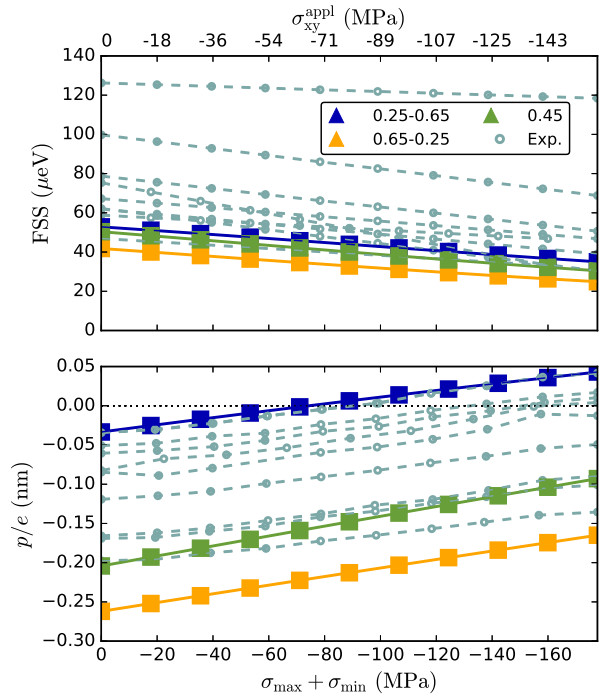} \\
\end{tabular}
\end{center}
\caption{
Dependencies of FSS (top panel) and $p/e$ (bottom panel) on $\sigma_{\mathrm{max}}+\sigma_{\mathrm{min}}$ experimentally obtained from $\mu$PL measurements of nine InGaAs QDs~\cite{Aberl:17} (broken curves) and that calculated for different In contents inside QD. The data for In content linearly varying as a function of vertical dimension from 0.25~(0.65) at the QD base to 0.65~(0.25) at the QD apex are shown as blue~(orange) curves. Those for constant In content of 0.45 are given as green curves. All other properties of the dots were the same as for QD$_2$ including $\sigma^\text{pre}=350$\,MPa.  The meaning of both x-scales is the same as in Fig.~\ref{fig:TheorVsExp}.
\label{fig:TuningByConc} 
}
\end{figure}

Motivated by Refs.~[\onlinecite{Fry:00}]~and~[\onlinecite{Grundmann:95}] which discussed the influence of indium distribution inside InGaAs/GaAs QDs on $p$, we have tested that observation for our stress-tuned dots. In Fig.~\ref{fig:TuningByConc} we show FSS and $p/e$ as a function of $\sigma_{\mathrm{max}}+\sigma_{\mathrm{min}}$ for In contents (i) linearly increasing from 0.25 at the QD base to 0.65 at its apex, (ii) the same but for reverted concentration profile and (iii) for constant In composition of 0.45. Similarly as in Refs.~[\onlinecite{Fry:00},\onlinecite{Grundmann:95}], we find that $p/e$ at $\sigma_{\mathrm{max}}+\sigma_{\mathrm{min}}=0$ can be varied considerably by changing the slope of In content from $-0.05$\,nm for (i) to $-0.27$\,nm for (ii). The case (iii) is found somewhat in between at $-0.21$\,nm. Note, that the calculated slopes $1/e\times\partial p/\partial\sigma^{\mathrm{appl}}_{xy}$
do not fit the experimentally observed ones so well as for different $\sigma^\text{pre}_{xy}$ discussed before.

On the other hand, the influence of different In gradients on
the values of FSS is much weaker than for $p/e$. This is expected since FSS is most sensitive to the in-plane QD symmetry \cite{Trotta:15} which is decreased in the presence of in-plane shear stress. Thus, the In gradient cannot be used to explain the spread of values of FSS that we have experimentally observed. Additionally, calculations for different QD height are shown in Ref.~[\onlinecite{SupMatHeight}].

\begin{figure}[!ht]
\begin{center}
\includegraphics[width=0.45\textwidth]{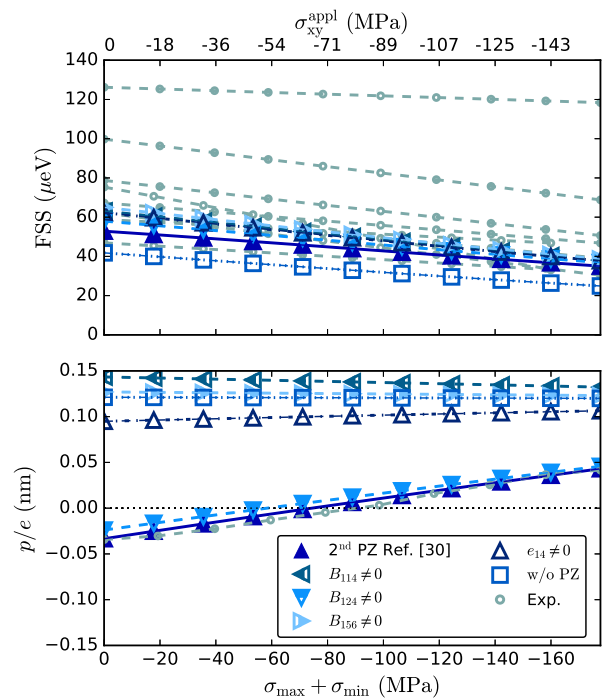}
\end{center}
\caption{
Comparison of dependencies of FSS and $p/e$ on $\sigma_{\mathrm{max}}+\sigma_{\mathrm{min}}$ for $\sigma_{xy}^\text{pre}=350$~MPa and all piezoelectric parameters equal to zero together with results for  $e_{14}$, $B_{114}$, $B_{124}$, and $B_{156}$ sequentially retaining their values for QD$_2$. For comparison, one set of the experimental data for $p/e$ from Ref.~[\onlinecite{Aberl:17}] is given by the grey broken curve in bottom panel. The meaning of both x-scales is the same as in Fig.~\ref{fig:TheorVsExp}.
\label{fig:DiffPiezoCoeff}}
\end{figure}

We now proceed with the analysis of the evolution of $p/e$ on $\sigma_{\mathrm{max}}+\sigma_{\mathrm{min}}$ and the apparent similarity of its slope among different QDs that we have measured. To investigate the origin of that we have performed calculations in which we have set all piezoelectric parameters equal to zero as well as sequentially $e_{14}$, $B_{114}$, $B_{124}$, and $B_{156}$ to the values listed in Tab.~\ref{Tab:PiezoConstants}, see Fig.~\ref{fig:DiffPiezoCoeff}.
\begin{table}
\caption{Used values in units of C/m$^2$ for the piezoelectric constants defined in Eqs.~(\ref{eq:1stPiez}, \ref{eq:2ndPiez})
as obtained from calculations given in Ref.~[\onlinecite{Beya-Wakata2011}].
For In$_x$Ga$_{1-x}$As, the constants were obtained by linear interpolation.}
\begin{ruledtabular}
\begin{tabular}{ccccc}
&$e_{14}$ & $B_{114}$ & $B_{124}$ & $B_{156}$ \\
\hline 
InAs & -0.115 &  -0.6 & -4.1 & 0.2 \\
GaAs & -0.238 & -0.4 & -3.8 &  -0.7 \\
\end{tabular}
\end{ruledtabular}
\label{Tab:PiezoConstants}
\end{table}

Firstly, by comparing the open squares with the full triangles in the top panel of Fig.~\ref{fig:DiffPiezoCoeff}, we note that FSS is dominated by $\sigma_{xy}^\text{appl}+\sigma_{xy}^\text{pre}$ and that the concomitant piezoelectric field ${\bf P}_{l}+{\bf P}_{nl}$ as given by Eqs.~(\ref{eq:1stPiez},\ref{eq:2ndPiez}) increases FSS by only $\sim 25\,\%$. As shown by the open triangles, this increase is overestimated twice by considering ${\bf P}_{l}$ only.  
Note that according to Eqs.~(\ref{eq:1stPiez},\ref{eq:2ndPiez}) the pre- and applied stress are in-plane and thus result in a purely perpendicular piezoelectric field. As shown in Ref.~[\onlinecite{Trotta:12_2}], electric fields in this direction couple to the FSS only via the different dipole moments of the respective excitons. Obviously, this coupling is less effective for FSS than the in-plane symmetry-breaking effect of  $\sigma_{xy}^\text{appl}+\sigma_{xy}^\text{pre}$. Moreover, the small response of FSS to electric fields in perpendicular direction justifies for our analysis of FSS \textit{a posteriori} the neglection of the n-i-p diode built-in electric field, which is estimated to be approximately two times smaller than the piezoelectric field.
%
%
%

Secondly, from the bottom panel of Fig.~\ref{fig:DiffPiezoCoeff} we see from the comparison of the effects of first-~and second-order piezo that the latter is dominant for $p/e$. In particular, the term containing the piezoelectric parameter $B_{124}$ in Eq.~(\ref{eq:2ndPiez}) almost exclusively determines the dependencies of $p/e$ on $\sigma_{\mathrm{max}}+\sigma_{\mathrm{min}}$. This is not surprising since the magnitude of $B_{124}$ is several times larger than that of $e_{14}$, $B_{114}$, or $B_{156}$.~\cite{Beya-Wakata2011} 
This observation, however, suggests a simplification of Eqs.~(\ref{eq:1stPiez})~and~(\ref{eq:2ndPiez}) by discarding all terms except for that for $B_{124}$. Let us now expand the $z$ element of ${\bf P}$ of the second term in Eq.~(\ref{eq:2ndPiez}) as
\begin{align}
\label{eq:2ndPiezModelPKexp}
P_z&=2B_{124}\eta_{xy}(\eta_{xx}+\eta_{yy})=\nonumber\\
&=2B_{124}\left(\eta^{\mathrm{QD}}_{xy}+\eta^{\mathrm{appl}}_{xy}+\eta^{\mathrm{pre}}_{xy}\right)\left(\eta^{\mathrm{QD}}_{H}+\eta^{\mathrm{appl}}_{H}+\eta^{\mathrm{pre}}_{H}\right),
\end{align}
where $\eta_{H}\equiv\eta_{xx}+\eta_{yy}$ corresponds to the hydrostatic in-plane strain. The meaning of the other variables is as follows: $\eta^{\mathrm{QD}}_{xy}$ is the shear strain stemming from the lattice mismatch between the dot material and GaAs matrix, $\eta^{\mathrm{appl}}_{xy}$ is the shear strain induced by the piezoelectric actuator, and $\eta^{\mathrm{pre}}_{xy}$ is the fixed shear prestrain; $\eta^{\mathrm{QD}}_{H}$, $\eta^{\mathrm{appl}}_{H}$, and $\eta^{\mathrm{pre}}_{H}$ denote the corresponding in-plane hydrostatic components. Variations of the strain fields over the QD volume are neglected, i.e,~all strain components represent values averaged over the QD volume. 
%
%

Since it is reasonable to expect that ${\eta^{\mathrm{appl}}_{H}, \eta^{\mathrm{pre}}_{H} \ll\eta^{\mathrm{QD}}_{H}}$, we can neglect $\eta^{\mathrm{appl}}_{H}$ and $\eta^{\mathrm{pre}}_{H}$ arriving at
\begin{equation}
\label{eq:2ndPiezModelPKstr}
{P_z\approx 2B_{124}\eta^{\mathrm{QD}}_{H}\left(\eta^{\mathrm{appl}}_{xy}+\eta^{\mathrm{pre}}_{xy} + \eta^{\mathrm{QD}}_{xy}\right)},
\end{equation}
%
%
%
shedding light to the reason why we see a linear dependence of $p/e$ on $\sigma^{\mathrm{appl}}_{xy}$ in our measurements.
%
%
In turn, in the presence of large hydrostatic strains typical for QDs lattice mismatched to the host crystal, Eq.~(\ref{eq:2ndPiezModelPKstr}) has to be used to calculate $P_z$ rather than the commonly used first order expansion given,~e.g.,~in Ref.~[\onlinecite{YuCardona}] which for our case would read
\begin{equation}
\label{eq:2ndPiezModelCardona}
P_z=2e_{14}\left(\eta^{\mathrm{appl}}_{xy}+\eta^{\mathrm{pre}}_{xy} + \eta^{\mathrm{QD}}_{xy}\right).
\end{equation}

We can now work out the approximate dependence of $p/e$ on $\sigma^{\mathrm{appl}}_{xy}$ as
\begin{equation}
\label{eq:DipoleApprox}
p/e\approx p_0/e+A^{\mathrm{QD}}\left(\sigma^{\mathrm{appl}}_{xy}+\sigma^{\mathrm{pre}}_{xy} + \sigma^{\mathrm{QD}}_{xy}\right),
\end{equation}
where $A^{\mathrm{QD}} = B_{124}C^{\mathrm{el}}\eta^{\mathrm{QD}}_{H}/eG$; $C^{\mathrm{el}}$ is a scaling factor that reflects the effect of quantum confinement on position of quasiparticles in QD and $G$ is the shear modulus. All built-in dipole moments independent of the piezoelectric polarization (induced,~e.g.,~by a gradient in the In concentration in QDs) are lumped together in $p_0$. 
%
%
%
%
According to Eq.~(\ref{eq:DipoleApprox}), only the off-diagonal element of  the prestress tensor is important for the simulation of $p/e$ in highly lattice mismatched QD systems, justifying the inclusion of prestress in our simulations by a single scalar parameter $\sigma_{xy}^\text{pre}$ as described in the beginning of the paper.     

\begin{figure}
\begin{center}
\includegraphics[width=0.42\textwidth]{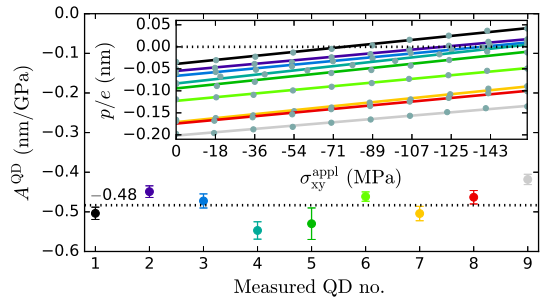} 
\end{center}
\caption{
Results of linear fits of experimental dependencies of $p/e$ on $\sigma_{xy}^\text{appl}$ by Eq.~(\ref{eq:DipoleApprox}). The colors of data points correspond to the colors of fitted linear lines in inset. The dotted curve corresponds to the mean value of $\overline{A}^{\mathrm{QD}}=-0.48$~nm/GPa.
\label{fig:DipoleModelTest}}
\end{figure}

In Fig.~\ref{fig:DipoleModelTest} we provide the test of Eq.~(\ref{eq:DipoleApprox}) by linear fitting of the experimental dependencies of $p/e$ on $\sigma_{xy}^\text{appl}$. It can be seen that the values for the slope $A^{\mathrm{QD}}$ for all studied QDs are scattered by less than $\pm 20\% $ around a mean value of $\overline{A}^{\mathrm{QD}}=-0.48$\,nm/GPa. Since $\eta_H^{QD}$ is the only experimental parameter in $A^\text{QD}$, we conclude that the uncertainty in the value for $A^\text{QD}$ is due to a variation of $\eta_H^{QD}$ of only $\pm 20\,\% $ for our QDs. Finally, the effective offset dipole moment given by  Eq.~(\ref{eq:DipoleApprox}) as $p_0^\text{eff}/e = p_0/e + A^\text{QD}(\sigma^{\mathrm{pre}}_{xy} + \sigma^{\mathrm{QD}}_{xy})$ pronouncedly varies by $\pm 75\,\%$ for the QDs shown in Fig.~\ref{fig:DipoleModelTest} as a consequence of variations of $p_0$ and $\sigma^{\mathrm{pre}}_{xy}$. To which extent each of them contributes to the observed variation of $p_0^\text{eff}$ cannot be concluded based on the experimental data available. Finally, by comparing the effects of Eqs.~(\ref{eq:2ndPiezModelPKstr})~and~(\ref{eq:2ndPiezModelCardona}) on $p/e$, respectively, using Eq.~(\ref{eq:DipoleApprox}) we find that the former provides $\approx 6$ times larger values of $A^{\mathrm{QD}}$ than the latter.
%
%


In conclusion, we have studied the effects of nonlinear piezoelectricity on
%
%
built-in electric dipole and excitonic fine-structure splitting energy in stress-tuned InGaAs/GaAs quantum dots and pinpointed its importance as compared to first-order terms only. Furthermore, it was found that while the dipole is influenced by the shear prestress via the piezoelectric effect, the latter effect is relatively unimportant for fine-structure splitting. On the contrary, shear prestress influences fine-structure by reducing the dot overall symmetry, particularly in the base plane of InGaAs/GaAs quantum dots.
Finally, we have found the dominant piezoelectric term and provided an approximate relation to estimate the influence of the applied stress on the electrical dipole moment for the InGaAs/GaAs QD system. The applicability of our simplified model extends also to other strongly lattice mismatched piezoelectric QD systems with large built-in hydrostatic strains. Its contribution to $P_z$  potentially dominates the more commonly used relation $P_z=2e_{14}\eta_{xy}$. Noticeably, in the case of InGaAs/GaAs quantum dots studied in this work our model led to almost an order of magnitude larger effect of applied shear stress on quantum dot dipole than that when only the linear piezoelectricity was considered.


P.K. would like to thank Prof.~Dieter Bimberg for fruitful discussions. A part of the work was carried out under the project CEITEC 2020 (LQ1601) with financial support from the Ministry of Education, Youth and Sports of the Czech Republic under the National Sustainability Programme II. The authors P.K., P.S., T.F., A.R., and R.T. were supported through the project MOBILITY jointly funded  by the Ministry of Education, Youth and Sports of the Czech Republic under code 7AMB17AT044 and by the Austrian Federal Ministry of Science, Research and Economy under the OEAD project CZ 07/2017. A.R. acknowledges financial support of the Austrian Science Fund (FWF): P 29603. Part of this work is financially supported by the European Research council (ERC) under the European Union’s Horizon 2020 Research and Innovation Programme (SPQRel, Grant Agreement No. 679183).


\bibliography{paper_jasz.bib}

\end{document}